\documentclass[superscriptaddress,aps,twocolumn,showpacs,showkeys,notitlepage,amsmath, amssymb,
]
{revtex4-1}

\usepackage{graphicx}
\usepackage[mathlines]{lineno}
\usepackage[svgnames]{xcolor}
\usepackage{hyperref}
\usepackage{color}
\hypersetup{
	colorlinks = true,
	linkcolor = Blue,
	citecolor = Blue,
	urlcolor  = Blue
}

\begin{document}

\title{Muon contact hyperfine field in metals: A DFT calculation}

\author{Ifeanyi John Onuorah}
\affiliation{Department of Mathematical, Physical and Computer Sciences, University of Parma, Italy}

\author{Pietro Bonf\`{a}}
\affiliation{CINECA, Casalecchio di Reno 6/3 40033 Bologna, Italy}
\author{Roberto De Renzi}
\email[]{roberto.derenzi@unipr.it}
\affiliation{Department of Mathematical, Physical and Computer Sciences, University of Parma, Italy}

\date{\today}

\begin{abstract}
In positive muon spin rotation and relaxation spectroscopy it is becoming nowadays customary to take advantage of Density Functional Theory (DFT) based computational methods to aid the experimental data analysis. DFT aided muon site determination is especially useful for measurements performed in magnetic materials, where large contact hyperfine interactions may arise. Here we present a systematic analysis of the accuracy of the \emph{ab initio} estimation of muon's hyperfine contact field on elemental transition metals, performing state of the art spin-polarized plane wave DFT and using the projector augmented pseudopotential approach, which allows to include the core state effects due to the spin ordering. We further validate this method in not-so-simple, non-centrosymmetric metallic compounds, presently of topical interest for their spiral magnetic structure giving rise to skyrmion phases, such as MnSi and MnGe. The calculated hyperfine fields agree with experimental values in all cases, provided the spontaneous spin magnetization of the metal is well reproduced within the approach. To overcome the known limits of the conventional mean field approximation of DFT on itinerant magnets, we adopt the so-called reduced Stoner theory [L. Ortenzi et al.,\href{https://journals.aps.org/prb/pdf/10.1103/PhysRevB.86.064437}{Phys. Rev. B 86, 064437 (2012)}]. We establish the accuracy of the estimated muon contact field in metallic compounds with DFT and our results show improved agreement with experiments compared to those of earlier publications.

\end{abstract}




\maketitle

\section{\label{sec:intro}Introduction}

Muon Spin Rotation spectroscopy ($\mu$SR) is widely employed to investigate new strongly correlated electron materials, whose spin and orbital correlations display interesting temperature behavior that may show up directly in the experimental muon decay anisotropy. A significant advancement in modeling of $\mu$SR data stems from the knowledge of the muon site, not granted a-priori and often provided by DFT calculations since the advent of High Performance Computing (HPC). It allowed, just to quote a few notable examples, the precise identification of specific muon bonds in insulators, \cite{bernardini2013,moller2013a} the identification of deep and shallow hydrogen states in semiconductors,~\cite{VandeWalle1990,Villao2011} the pressure induced magnetic structure in MnP ~\cite{Khasanov2016} and the determination of infrequent subtle muon induced effects in rare-earth pyrochlores.~\cite{Foronda2015}

However, the crucial point that provides {\em quantitative} access to electronic spin degrees of freedom is the full knowledge of the muon couplings with its surroundings. The often missing key ingredient is the contact hyperfine interaction, notably relevant in metals. This quantity may be calculated by ab-initio techniques but in practice the few published results date back to the early developments of DFT.

Only the determination of the muon implantation site {\em and} of the interaction constants between the muon and its atomic surrounding give access to crucial material properties such as the value of the ground state ion magnetic moment in ordered materials and possibly the magnetic structure as well. Although $\mu$SR cannot compete with diffraction techniques for magnetic structure determination, there are cases where the latter are not applicable (due to the presence of either strong incoherent scatterers or neutron absorbers)~\cite{Pascua2014} or not sufficient for a complete determination.~\cite{Khasanov2016} A noteworthy example is provided by the recent refinement of an additional structural parameter in the zero-field cycloid state of MnSi and its skyrmion phase, a non vanishing phase between the two Mn orbits in the cycloid,~\cite{Dalmas2016,Dalmas2017} inaccessible to neutrons, whose determination by $\mu$SR is made possible by the low symmetry of the muon site. The knowledge of both site and contact couplings are essential for this information to be retrieved.  

Here we provide a demonstration of the effectiveness of a DFT-based approach, validated by the comparison with available experimental determinations.  Five materials are selected by this criterion from the literature, ranging from simple magnetic metals, Fe, Co, Ni, to two additional chiral magnets of current high interest, MnSi and MnGe. The list of metals where the hyperfine coupling is experimentally known is unfortunately scarce, since they require quite accurate and time consuming experiments on single crystals, and this is actually an additional motivation for validating a more general ab-initio method.

The structure of the paper is the following: Sec.~\ref{sec:musr} briefly reviews the experimental technique highlighting the requirements for the theoretical approaches together with the most significant recent improvements; in Sec.~(\ref{sec:comptdet}) we analyze the different computational aspects that allow to obtain converged results; in Sec.~\ref{sec:discussion} we discuss our results distinguishing the three elemental transition metals, Fe, Ni and Co, used as a  benchmark of the theoretical approximations from the case of the two additional metallic materials of current interest, MnSi and MnGe. 

  Finally, we discuss additional possible refinements to further reduce the difference between calculated and measured values.

\section{\label{sec:musr}$\mu$SR and the muon couplings}

$\mu$SR exploits the implantation of spin polarized muons to probe local properties of materials, by means of the local magnetic field at the muon, together with its dynamics on the scale of the muon's mean lifetime ($\approx$ 2.2 $\mu$s). Notably, this experimental technique makes predominant use of the positive antiparticles ($\mu^+$). The basis of this technique lies in the anisotropic positron emission at the muon decay, peaked around the muon spin direction. The anisotropy is a hallmark of weak interactions in the three-body decay, and the very high muon spin polarization (almost 100\%) relies on the same violation in the two-body pion decay that originates this probe particle. The evolution of the muon spin direction may be thus detected over several microseconds, with very fine time resolution, over an ensemble of individually implanted particles. Thus, $\mu$SR may be considered akin to Nuclear Magnetic Resonance (NMR), with the advantage of a broader applicability and a non resonant broadband detection.  The foremost applications are in superconducting and magnetic materials, including very weak magnets, thanks to the good sensitivity provided by the large muon gyromagnetic ratio ($\approx$ 135 MHz/T).~\cite{schenck1995,yaouanc2011} 

Implanted muons thermalize in inorganic crystalline solids almost invariably at interstitial sites in the lattice, so that the  detected internal field is that at an interstitial, extremely diluted impurity. The experimental value of the muon local field, including both its static value and its fluctuating dynamical components, provides important clues towards understanding the magnetic properties of the host material. The muon local field both in superconductors and in magnetically ordered materials yields the temperature dependence of the order parameter, critical fluctuations are reflected in relaxation rates, the presence of additional phase transitions is easily detected, just to quote a few examples. All these properties are directly accessed from the $\mu$SR spectra without any prior knowledge of the muon site and the details of its couplings. 

In the following we shall refer explicitly to the investigation of magnetic materials, a specialty of $\mu$SR. Typically, here any refinement of the analysis does require additional a priori knowledge of the muon implantation site. 

Ab-initio Density Functional Theory (DFT) prediction of the muon site has been successfully employed in several studies, starting from the early pioneering investigations to the present more extensive, although not yet every-day use, as detailed in a few reviews on the subject.~\cite{moller2013b,bonfa2016,bernardini2013} Site assignment is the key initial ingredient in the non infrequent cases where the internal magnetic field is dominated by the distant dipole contribution, that requires only the knowledge of the site in order to be computed by a classical sum over the dipole moments of the host lattice.~\cite{schenck1995, yaouanc2011,bonfa2018} Thus, the comparison between predicted and measured local field can validate the muon site assignment, and in turn, this assessment yields, e.g., a measure of the magnetic moment values. However, additional local field contributions exist and they are not negligible in many cases. Thus, a non trivial ab-initio calculation of the couplings, besides its intrinsic value, in some cases are crucial for the site validation itself.

The contributions to the experimental local field, besides the already mentioned dominant dipolar sums, include another trivial term that is shape-dependent (demagnetization) and proportional to the macroscopic sample magnetization.~\cite{yaouanc2011} We shall concentrate here on the contributions that require a quantum mechanical description of the host electrons in the vicinity of the probe. In a localized spin magnet, they may give rise to direct transferred and super-transferred couplings, depending on whether the wave-function overlap between the muon probe and the magnetic ion is direct or through the polarization of the wave-functions of intervening ligands. In metals, the  conduction electrons provide an example of the first kind, giving rise to a contact interaction term, that results in a spin density at the muon site. For the purpose of this paper we will focus only on the contact hyperfine interaction at the muon.  

In the absence of external magnetic field and within a non relativistic quantum mechanical description, the local field resulting from the interaction between the muon and an $s$-electron at distance $r_{e-\mu} \to 0$ from the muon is described by the following Hamiltonian~\cite{slichter1996principles}

\begin{align}
\mathcal{H} = \frac {2\mu_0} {3} \gamma_e \gamma_\mu \hbar^2 \mathbf{S}_\mu  \mathbf{S}_e \delta(\mathbf{r})
\label{eq:equation1}
\end{align}
where $\mu_0$ is the vacuum permeability,  $\gamma_e$ and $\gamma_\mu$ are the electron and muon gyromagnetic ratio respectively while $\mathbf{S}_e$  and  $\mathbf{S}_\mu$ are their spin operators. It has been assumed that the muon is point-like. In the collinear spin approximation, by integrating over the electron coordinates, the contact hyperfine field at the muon ${B}_c$ is \cite{giustino2014materials}
\begin{align}
{B}_c = \frac {2} {3} \mu_0 \mu_B \rho_s (\mathbf{r}_\mu)
\label{eq:equation2}
\end{align}
where  $\mu_{B}$ is the Bohr magneton, and the spin density, $\rho_s $  defined as $ (\rho _{\uparrow}(\mathbf{r}_\mu) -\rho_{\downarrow}(\mathbf{r}_\mu))$  with $\rho _{\uparrow}$ and $\rho _{\downarrow}$ being the density associated to each spinor component at the muon site  $\mathbf{r}_\mu$.
This equation was used to evaluate the contact field at the muon with the spin polarization obtained from DFT simulations.

First principle theory of the hyperfine parameters for both heavy and light nuclei in magnetic materials is in principle well understood and has been studied back from the mid 1960's.~\cite{daniel1963, jena1979,meier1975,rath1979,patterson1979,jena1981,estreicher1981,lindgren1982,terakura1983,blugel1987,walle1993} Various approaches were proposed to improve the accuracy of the calculated contact fields, but these investigations, in particular for the muon in metals, were undertaken when computing resources were orders of magnitude less powerful than today. Their results are compared with our calculations in Sec.~\ref{sec:discussion}. More recently valuable theoretical improvements~\cite{mauri1996a,pickard2001,avezac2007,laskowski2015,nusair1981,pavarini2001} have established DFT as the standard for the calculation of NMR shift parameters, most reliably in non magnetic insulators. However, these improved methods were never directly applied to the muon case in metals. The main difference as already noted, is that the location of the nuclei is extremely well known from diffraction, whereas the determination of the muon site is part of the same DFT problem, requiring in addition large supercells to represent the ideally diluted impurity while keeping an accurate description of the bulk sample. With the current availability of HPC it is well due to extend these modern methods to muon studies in metallic systems in order to establish their accuracy and applicability. 

\begin{table*}
\caption{\label{tab:table1} Muon sites in fractional coordinates; spin only magnetic moment for each magnetic ion for the conventional GGA calculation $m_{GGA}$,  the experimental value $m_{exp}$ and the reduced Stoner theory calculation (see Sec.(~\ref{sec:spyral}) ) $m_{RST}$ in units of $\mu_B$; calculated spin density at the muon in atomic units of $(a_0^{-3})$ resulting from the pseudo-wavefunction $\rho^{PS}_{s}$ and the PAW reconstructed value $\rho^{PR}_{s}$.} 

\begin{ruledtabular}
 \begin{tabular}{ l c c c c c c c }
    Host metals~\footnote{The MnGe and MnSi structure are of P$2_1$3 space group (cubic) with the Mn atom at (0.138,0.138,0.138) crystal unit position.} & $ Muon$ $sites$ & $m_{GGA}$  &  $m_{exp}~\footnote{See Refs.~\cite{kittel1996,martin2016,amato2014}} $& $m_{RST}$ & $\rho^{PS}_{s} (\mathbf{r}_\mu)$  &  $  \rho^{PR}_{s} (\mathbf{r}_\mu)$  \\ 
     \hline
    Fe - bcc&0.50, 0.25, 0.00 &2.17 & 2.22 & - &  -0.0179 &-0.0238 \\ 
    Co - hcp &0.33, 0.67, 0.25& 1.585& 1.72 &- & -0.0111 &-0.0150\\
    Co - fcc &0.50, 0.50, 0.50& 1.645 & 1.59 & - & -0.0109 &-0.0139\\
    Ni - fcc &0.50, 0.50, 0.50& 0.638 & 0.606 & - &  -0.0020 & -0.0028\\ 
    MnGe&0.552, 0.552, 0.552 & 2.014 & 1.83  & 1.84 &-0.0162 & -0.0217 \\
    MnSi&0.541, 0.541, 0.541&1.00 & 0.4  & 0.401 & -0.0031& -0.0042  \\ 
    \end{tabular}
\end{ruledtabular}
\end{table*}

\subsection{\label{sec:comptdet}Calculation details}

The pseudopotential and plane-waves (PW) basis approach as in the Quantum ESPRESSO suite of codes were used for our calculations.~\cite{qe2009} PW based codes have a number of important features, namely good parallel performances, good accuracy for the description of the bulk material and simplicity of the basis set.  The plane-wave basis is generally used to describe artificially smooth pseudo-wavefunctions thus avoiding the strong oscillations in the core region. Nonetheless, the Projector Augmented-Wave (PAW) method introduced by Bl\'ochl ~\cite{walle1989,walle1990,walle1993,blochl1994} allows to approximate the all-electron density using a frozen-core reconstruction starting from the pseudo wavefunction. In the context of the PW basis, the PAW reconstruction method is therefore the method of choice for an accurate evaluation of Eq.~\ref{eq:equation2}. Since periodic boundary conditions are implied in the description of the bulk system, the effect of the extremely diluted muons in the material must be modeled within the supercell approximation which reduces the artificial interactions between the charged impurities. It must be carefully verified that these artificial interactions of the muon periodic images are negligible on the quantities under study.

\begin{figure}
\centering
\includegraphics[width=8.5cm]{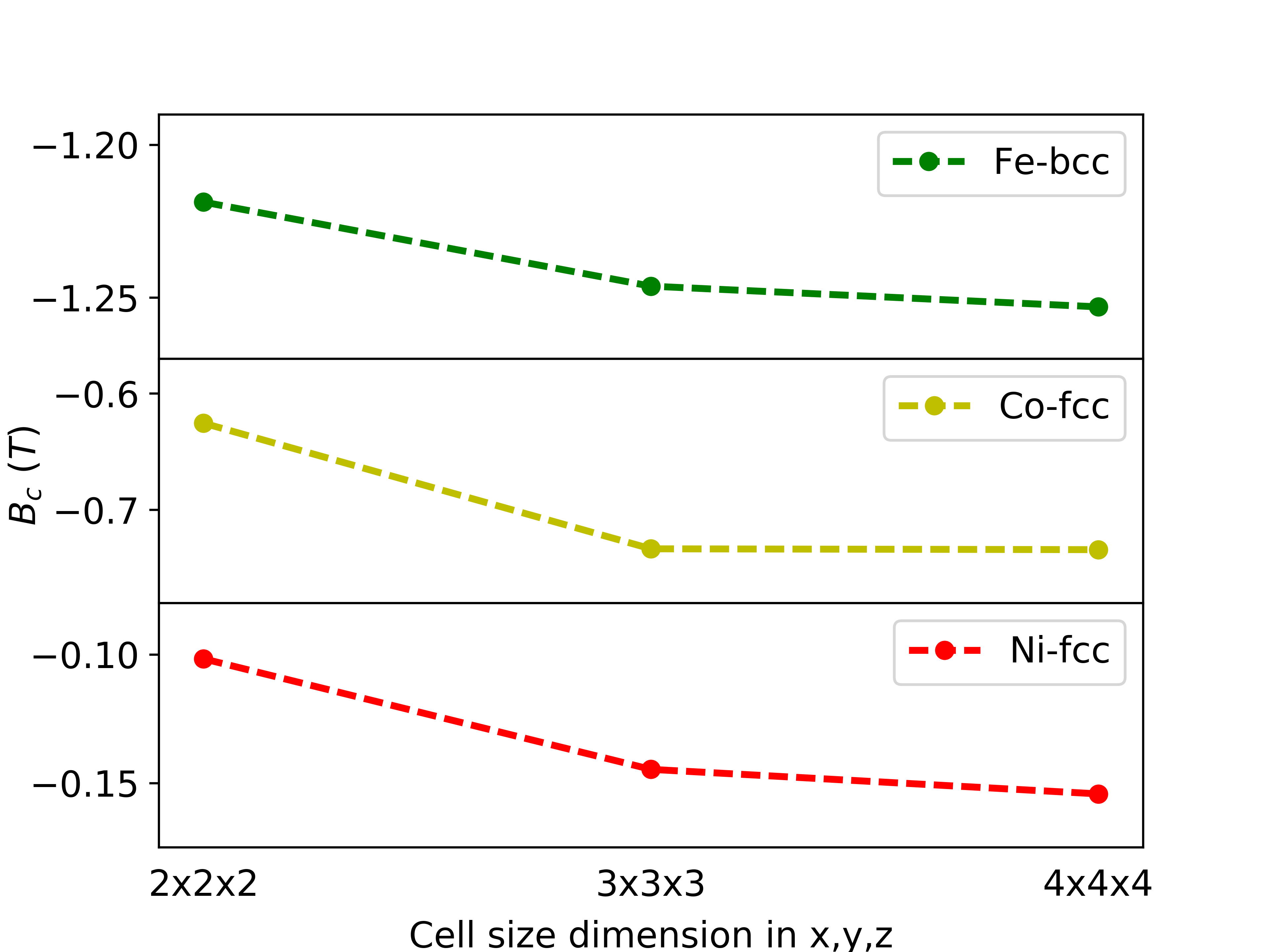}
\caption{\label{fig:figure1} Convergence of the muon contact hyperfine field $B_c$ with the size of the cell for host systems of  Fe, Ni and Co.}
\end{figure}

For all the calculations in this work, the plane wave cutoffs were always above 100 Ry, granting a convergence on total energy (threshold $10^{-4}$ Ry) and spin density, while the exchange correlation functionals were treated within the semi-local Generalized Gradient Approximation (GGA) using the Perdew-Burke-Ernzerhof (PBE) formalism.~\cite{pbe1996} The calculations were done in the scalar relativistic approach, neglecting spin-orbit coupling. The scalar relativistic approximation is sufficient for the theoretical calculation of the muon contact field since hydrogen (hence the muon) has a small nuclear charge and the contact field is predominantly due to on-site contributions of s-like states surrounding the muon.~\cite{blugel1987,battocletti1996} 
The Marzari-Vanderbilt~\cite{mv1999} smearing function was used.

\begin{figure}
\centering
\includegraphics[width=8.2cm]{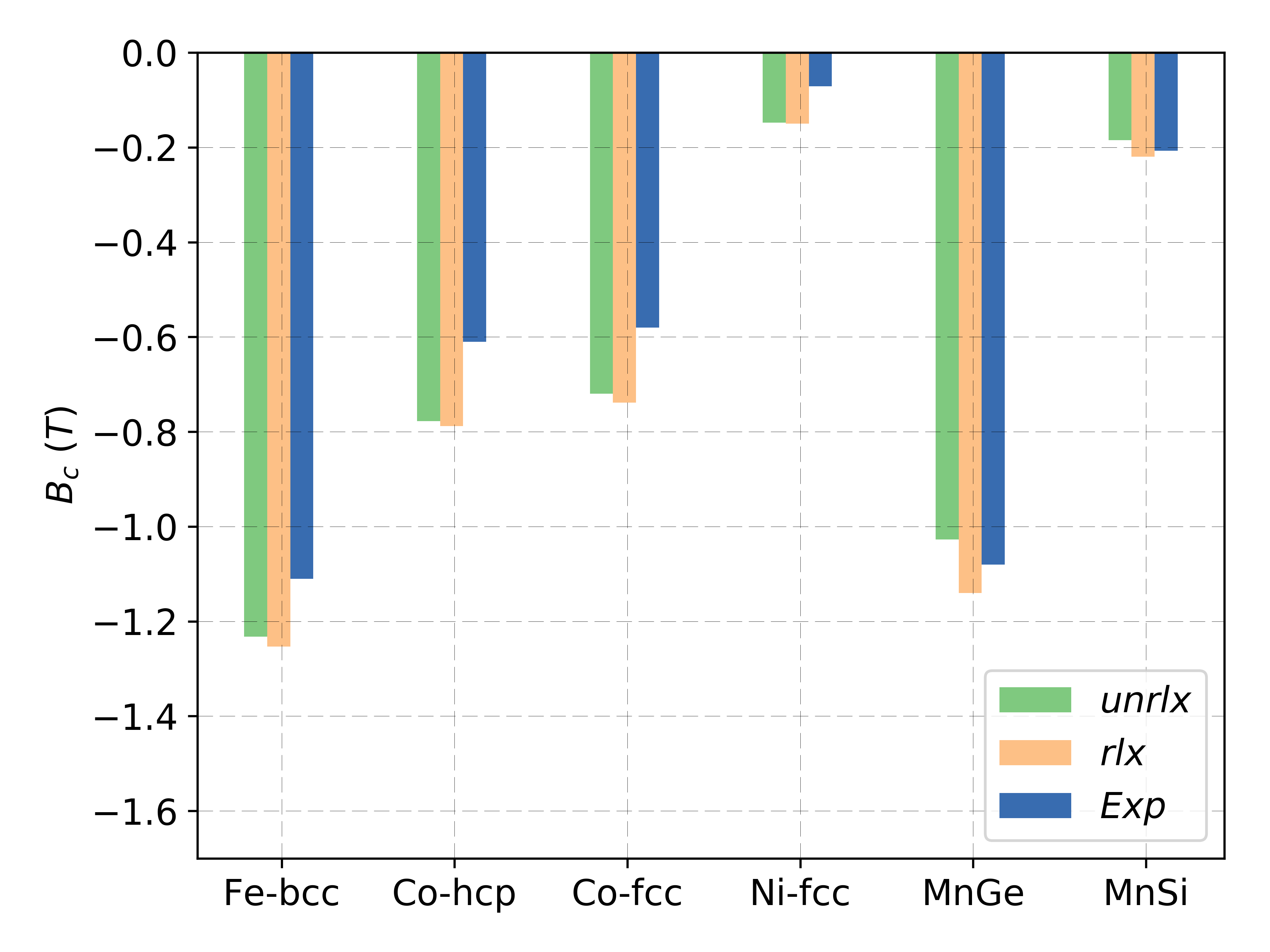}
\caption{\label{fig:figure2}Calculated contact hyperfine field $B_c$ for unrelaxed ($unrlx$) and relaxed host atoms + muon at fixed cell volume ($rlx$), compared to the experimental value ($Exp$).}
\end{figure}  
  
A uniform Monkhorst-pack~\cite{kpoint1976} mesh was used for the k-points. The convergence of the contact field depends strongly on how dense is the mesh of k-points. A 16 x 16 x 16 mesh grid was used for the unit-cell of the transition metals  and a 12 x 12 x 12 grid for the unit-cell of the B20 compounds. The mesh size were selected following a systematic test to ensure independence on the size used to the spin density and total energies. These grids were down-scaled proportionally for each supercell size. 

The first step for all calculations involves the optimization of the structure and the correct reproduction of the electronic and magnetic properties of the pristine material. The next step involves investigating the extent of the lattice distortion around the muon and its influence on the electronic and magnetic properties of the nearest neighbors. 
Before systematically comparing  calculations with experimental values, let us further notice that we expect our results to overestimate the experimental absolute value, in view of the light mass of the muon, which results in relatively large amplitude of zero point vibrations.~\footnote{the quantum average of the spin density at the muon results in a {\em reduction} of the point value, since the spin density starts to decay away from that point.} The muon behaves as a quantum oscillator and the extent of its wave function is completely neglected in the {\em static} contact field from the DFT approach. The experimental value should be compared to the average over the muon wave function, whose accurate determination will be addressed separately and is beyond the scope of the present paper.
\begin{figure}
\includegraphics[width=8.3cm]{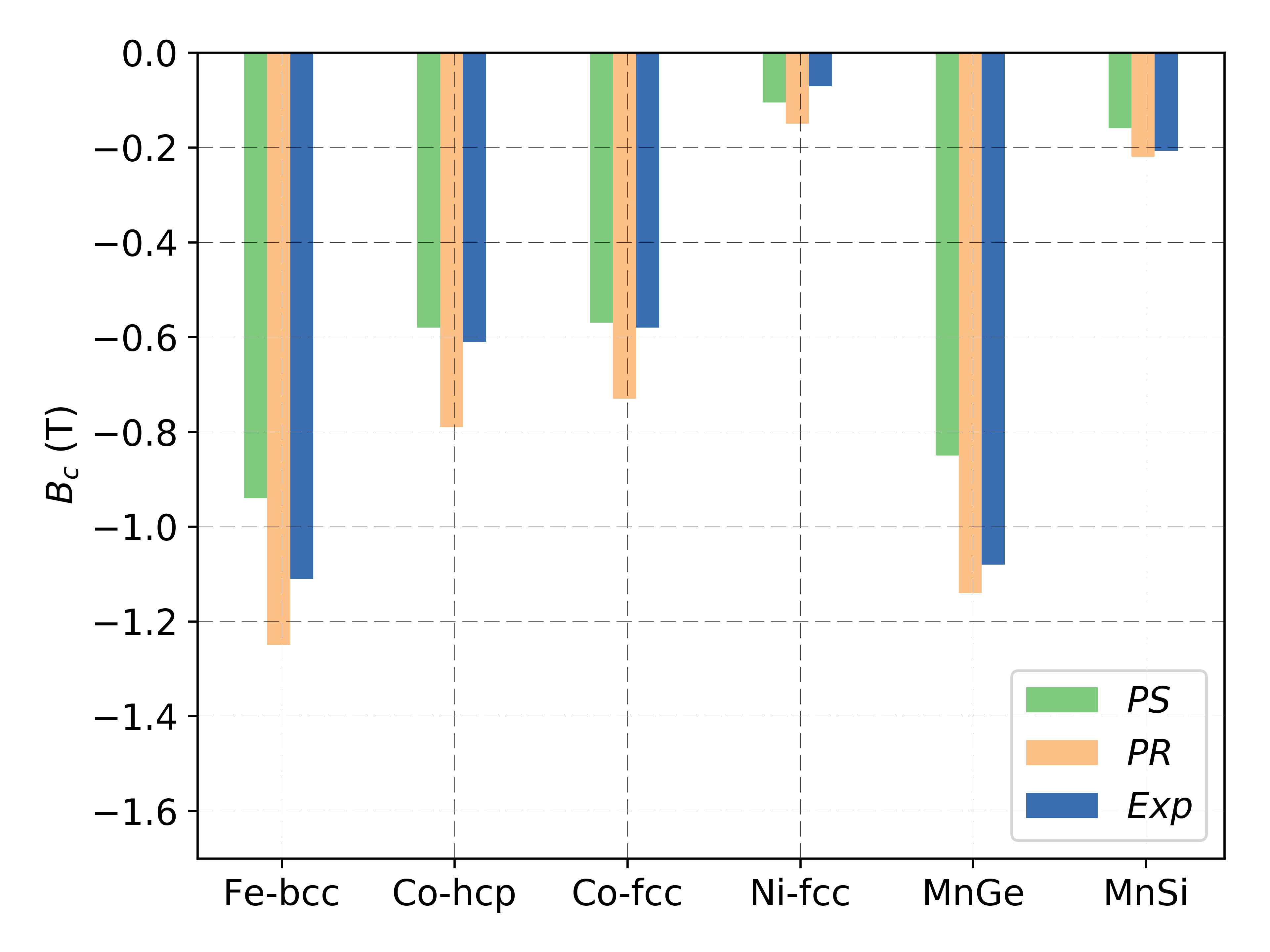}
\centering
\caption{\label{fig:figure3} Contact hyperfine field $B_c$ calculated with the spin densities from the pseudo-wavefunction ($PS$) and from the all electron reconstruction with the PAW method ($PR$), compared to the experimental value ($Exp$).}

\end{figure}

\section{\label{sec:discussion}Results and Discussion}
The appropriate size of the supercell for each of the materials was carefully determined considering convergence of the total DFT energies, distortion of the lattice and magnetic coupling in the vicinity of the muon  and in particular the calculated contact hyperfine field, as shown in Fig.~\ref{fig:figure1}. The plot shows that this quantity converges at the 3x3x3 cell level, however we will compare results on the transition metals obtained with 4x4x4 cells. Following the same systematic tests, a 2x2x2 cell was used for the B20 compounds. Incidentally, in these metals convergence is achieved when the muon periodic replica are above 8.48 \AA\ apart.

Next, we address the issue of whether the relaxation of the host atoms in the supercell including the muon has a significant effect on the quantity of interest. This is obtained comparing muon contact field values $B_c$  with and without atomic relaxation. The results indicate that the relaxation around the muon affects significantly only the positions of the nearest neighbor ions. The distortions are short ranged and small because the positive charge of the muon is screened by the electron cloud in metals, a fact that is directly shown in Fig.~\ref{fig:figure4} (see Appendix). Furthermore, Fig.~\ref{fig:figure2} shows that the direct effect of relaxation on the value of $B_c$ is tiny compared to the deviation between experiment and theory at this level of approximation. However, the reported results in this work are those of the relaxed lattice.  

Finally, we want to determine the relative accuracy of the pseudo-wavefunction ($PS$) spin densities compared to those obtained by the PAW Reconstruction ($PR$) method described in Sec.~\ref{sec:comptdet}. These are reported in the last two columns of Table~\ref{tab:table1} as $\rho^{PS}_{s}  (\mathbf{r}_\mu)$ and $\rho^{PR}_{s}  (\mathbf{r}_\mu)$ respectively and the corresponding contact field is plotted in Fig.~\ref{fig:figure3} and compared with the experimental values. The pseudo wavefunctions eventhough do not include the actual core electron density, give results remarkably close to the experimental values. It should be noted however that this is probably due to an error compensation between the approximated core electronic density and the missing zero point vibration corrections. In addition, the overshooting of all estimations obtained with PR is in agreement with the fact that corrections due to zero point vibrations may lead to a reduction of the absolute value (as mentioned in Sec.~\ref{sec:comptdet}) thus, systematically improving the agreement with the experimental data for all the compounds reported in Fig.~\ref{fig:figure3}.

\subsection{\label{sec:transition metals}Fe, Co, Ni }

We have investigated the accuracy of the calculated magnetic moments and the effects of the muon on them. The experimental total magnetic moments of the transition 3d metals, shown in Table~\ref{tab:table1}, are well reproduced within the conventional GGA-DFT. The tabulated magnetic moment were estimated  with the L\"{o}wdin population analysis.~\cite{sanchez1995} With the muon impurity in the lattice, the moment of the nearest neighbor host magnetic ions are negligibly perturbed. These perturbations contribute to no appreciable change of the calculated contact field. As we will further discuss, the contact field depends strongly on the accuracy of the calculated spin moments.   

\begin{table} [!tbp]
\caption{\label{tab:table2}  Calculated static contact hyperfine field at the muon $B_c$ by PAW reconstruction together with results from other works, experimental values $B_c^{exp}$ and deviations $\Delta^{exp} = B_c^{exp}-B_c$.}

 \begin{tabular}{ c @{\hspace{1em}} c @{\hspace{0.8em}} c @{\hspace{0.8em}} c @{\extracolsep{0.3cm}}  c}
    \toprule
     & \multicolumn{3}{c}{$B_c$ [T]}  \\
     \cline{2-4}
   Host metals & this work  & other works & exp & $~\Delta^{exp} [T]$ \\ 
     \colrule
    Fe-bcc  &-1.25   & -0.94~\cite{jena1979}    & -1.11~\cite{nishida1977} & 0.14 \\
      ,,   &        & -1.01~\cite{keller1979} \\
       ,,   &        & -1.44~\cite{lindgren1982} \\
       ,,   &        & -1.03~\cite{terakura1983} \\
     Co-hcp  &-0.79  & -1.34~\cite{jena1979} & -0.61~\cite{graf1976} & 0.18\\ 
        ,,   &       & -0.57~\cite{keller1979}\\
    Co-fcc   &-0.73  & -0.46~\cite{lindgren1982} & -0.58~\cite{lindgren1982} & 0.15\\ 
    Ni-fcc   & -0.15 & -0.69~\cite{jena1979}& -0.071~\cite{graf1979} & 0.08\\ 
       ,,    &       & -0.059~\cite{keller1979}\\
       ,,    &       &  -0.13~\cite{lindgren1982}\\
       ,,    &       & -0.31~\cite{rath1979}\\
       ,,    &       & -0.059\cite{patterson1979}\\
    MnGe & -1.14 & -  & -1.08~\cite{martin2016} & 0.06\\
    MnSi & -0.22 & -  &  -0.207~\cite{amato2014} & 0.013\\ 
    \botrule
    \end{tabular}   
\end{table}

The first important result obtained is that the calculated spin imbalance at the muon, shown in Table~\ref{tab:table1}, is negative for all the considered metals, in agreement with experiment and with the simple notion that the majority spin electrons are in a direction opposite to the bulk magnetization at the muon. Furthermore, the deviations, reported in  Table~\ref{tab:table2}, are on average 0.14 T and always within 0.2 T. This may be considered a rather good agreement, compared to results from the earlier works, since the averaging due to the muon's vibration is not included yet. 

Admittedly, many of these earlier works~\cite{jena1979,rath1979,jena1981,estreicher1982} estimated the spin density at the muon site by a simple re-scaling of the spin density of the bulk material at the position of the known muon site with an empirical spin enhancement factor that mimics the perturbation induced by the interstitial muon. This is clearly an unpractical ad-hoc solution that impairs the ab-initio method. They thus failed to establish the accuracy of the method over several materials.

In the earlier calculations, the large deviations between calculated and experimental contact field values (on average) were consistently attributed to the lack of muon zero point motion correction. Our more accurate results indicate that the effect of the zero point motion is needed but its extent is much smaller.

\subsection{\label{sec:spyral}MnGe and MnSi}

The muon implantation sites for MnSi and MnGe~\cite{amato2014,martin2016} are reported in Tab.~\ref{tab:table1}. Their zero field magnetic structure, actually a spin spiral, was approximated by a collinear ferromagnetic state since in both cases the pitch~\cite{deutsch2014,ishikawa1976,amato2014,makarova2012,deutsch2014,martin2016} is much longer than the lattice parameter. 
 
 The conventional DFT calculated spin only moment, $m_{GGA}$, deviates significantly from the experimental total magnetic moment for both B20 compounds, and for MnSi in particular. This is a consequence of the poor standard DFT description of  spin fluctuations in the magnetic ground state especially for itinerant electron systems. This also affects the calculated spin-density at the muon. For MnSi $m_{GGA}=1.0 \mu_B$, while the experimental value is $m_{exp}= 0.4 \mu_B$. Notably, the ratio of these two values  matches the ratio of the calculated  and experimental contact fields, for the calculated spin density of -0.0107 ($a_0^{-3}$). This is also the case for MnGe (see Table~\ref{tab:table1}), with calculated spin density -0.0251 ($a_0^{-3}$). Thus, the accuracy of the calculated contact field is heavily influenced by how well the host ground state magnetization is reproduced by DFT. A simple but non ab-initio way to predict experimental contact field values would consist in re-scaling the fields by the ratio $m_{exp}/m_{GGA}$ or constraining the total moment of the bulk material~\cite{PhysRevB.70.075114} to the known experimental value.
 
Ab-initio approaches have been discussed in the literature for MnSi.  Attempts to obtain the experimental local moment by the reduction of the lattice constant within the local density approximation (LDA)~\cite{carbone2006} work only for unphysical lattice constant values.  Hubbard U correction (DFT+U) to redistribute electrons between the majority and minority channels~\cite{collyer2008,shanavas2016} acknowledge unphysical results in the pressure dependence of the magnetic moment (and we checked that the spin density at the muon departs from experiment). 
 
A different approach was proposed by Ortenzi~\cite{ortenzi2012} who implemented a reduced Stoner theory (RST) modification to the exchange-correlation functionals. This approach involves the reduction of the ab-initio Stoner parameter in the conventional spin polarized DFT, by a spin-scaling factor (ssxc) in the exchange correlation potential.  The energy gain due to spin polarization is shown to be reduced as according to Moriya's self-consistent renormalization (SCR) theory.~\cite{moriya1973} The SCR theory is known to describe successfully the ground state properties of weak itinerant ferromagnet, and in particular that of MnSi.~\cite{moriya1984,moriya1985}

This method is variational and it adjusts the magnitude of the spin polarisation for all standard functionals. We re-implemented it in the Quantum Espresso code, and obtained $m_{RST}\approx m_{exp}$ with ssxc values of 0.83 and 0.95 respectively for MnSi and MnGe, as in Table~\ref{tab:table1}. The band structure remains negligibly changed, although band are shifted in energy accordingly to the reduced Stoner parameter. Our results for the contact field, in good agreement with experiments, were obtained with spin densities calculated from this approach and are summarized in Table~\ref{tab:table2}.

\section{\label{sec:conclusion}Conclusions}

 We have reviewed and validated a systematic approach to the calculation of the muon's static contact hyperfine field in metals, to aid in ${\mu}SR$ data analysis and understand the contact contribution to the hyperfine field of light impurities in metals. 

We have successfully established the accuracy of the estimation of the muon contact field in metallic compounds with DFT. 
The pseudopotential DFT approach within the PAW formalism is good even for itinerant magnets, notoriously difficult systems. 

The results may be affected by poor DFT reproduction of the magnetic moment which is common for these systems. The RST method allows a variational approach that may well reproduce the experimental results (both the magnetic moment and spin density) without forcing unphysical values of the other lattice quantities. 

Our final results are obtained for an infinite muon mass and do not account for its finite zero point vibrations. The full treatment of this aspect is beyond the scope of the present work, but we know that it has to reduce the absolute value of the static spin density. Therefore, the fact that the calculated value $|B_c^{exp}|-|B_c|$ is consistently small and negative agrees with the expected effect of zero point vibrations. With this in mind the agreement between calculation and experiment is satisfactory at this level of approximation.

We conclude that DFT calculation of contact hyperfine fields is a viable assistance to $\mu$SR data analysis. 
Its standard implementation may well replace expensive, time consuming and often not readily available  single crystal measurements in the future.

\begin{acknowledgments}
We thank Luciano Ortenzi for useful discussions. The calculations were performed with computing resources provided by STFC Scientific Computing Department’s SCARF cluster, the Swiss National Supercomputing Centre (CSCS) under project ID sm07 and the hpc resources at the University of Parma, Italy. We also acknowledge grants from the European Union's Horizon 2020 research and innovation program under grant agreement No. 654000.

\end{acknowledgments}

\appendix*
\section{\label{sec:appendix} Short range distortion due to the muon}

\begin{figure}[h!]
\includegraphics[width=9.0cm]{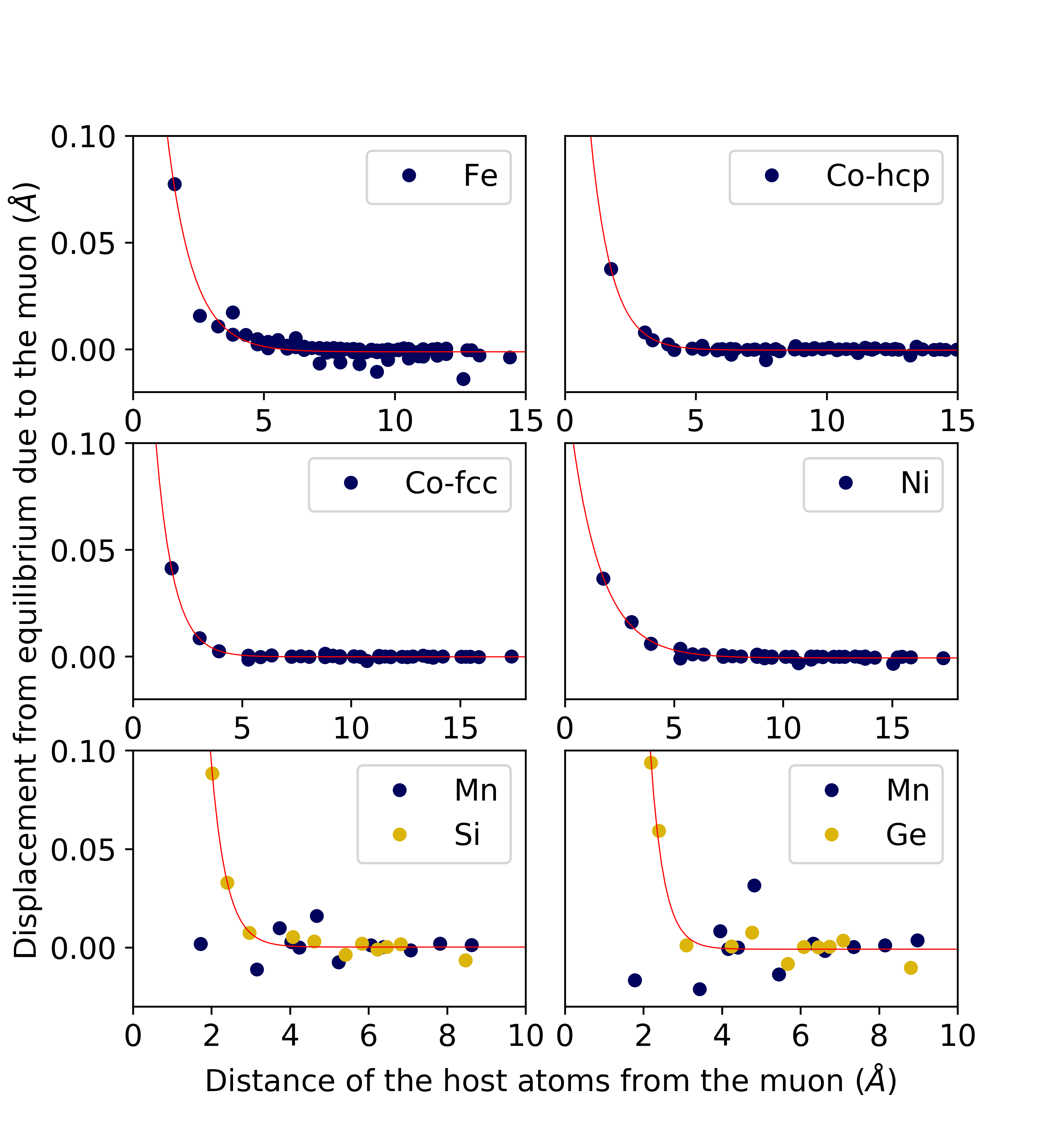}
\centering
\caption{\label{fig:figure4} Atomic displacement in the presence of the muon in a relaxed supercell, vs. the distance of each atom from the muon. Red lines are guide to the eye showing the exponential decay.}
\end{figure}

The range of the lattice strain introduced by the interstitial muon defect may be directly quantified by comparing the position of each atom in the pristine material with their position in the supercell DFT calculation, after lattice relaxation with the muon. Fig.~\ref{fig:figure4} shows the difference of these two quantities versus the distance from the muon site. The top four panel display the result for the 4x4x4 cell of the elemental metals, with a clear exponential decay on a length-scale, $\lambda < 1.25 $ \AA.

The data for the 2x2x2 cell of the B20 compounds are more scattered, as expected in view of the presence of two different species. Interestingly, Si and Ge show a decaying displacement with $\lambda < 3.0$ \AA\ whereas Mn ions show no systematic deviation, perhaps indicating a much shorter value of $\lambda$ . 

\bibliography{references}

\end{document}